\documentclass[twocolumn,twoside]{revtex4}
\usepackage{graphicx}
\usepackage{fancyhdr}
\usepackage{subfigure}
\usepackage{color}
\usepackage{lipsum}

\newcommand{\newc}{\newcommand*}
\newc{\gev}{\ensuremath{\,\mathrm{GeV}}}
\newc{\tev}{\ensuremath{\,\mathrm{TeV}}}

\newc{\be}{\begin{equation}}
\newc{\ee}{\end{equation}}
\newc{\bea}{\begin{eqnarray}}
\newc{\eea}{\end{eqnarray}}

\definecolor{LightGreen}{rgb}{0.88, 1, 0.88}
\begin{document}

\title{Global Bayesian Analysis of new physics in $b \to s \mu\mu$ transitions after Moriond-2019}

\author{Dinesh Kumar, Kamila Kowalska and Enrico Maria Sessolo}
\affiliation{National Centre for Nuclear Research, Pasteura 7, 02-093 Warsaw, Poland}

\begin{abstract}
The recent measurement of $R_K$ at LHCb continues to support the hint of violation of lepton flavor universality. We perform a global fit for new physics
in semileptonic $b\to s$ transitions using all the relevant data with a Bayesian
analysis technique. We include new measurements of $R_K$ at LHCb and new determinations of 
$R_{K^*}$ and $R_{K^{*+}}$ at Belle. We perform the scan for various NP scenarios and infer the 68\% and 95.4\% credibility regions of 
the marginalized posterior probability density for all scenarios. We also compare the models in pairs by calculating the Bayes factor 
given a common data set. A few well-known BSM models are analyzed that can provide a high energy framework for the EFT analysis.
These include the exchange of a heavy $Z^{'}$ boson in models with heavy vector-like fermions and a scalar field,
and a model with scalar leptoquarks. We provide predictions for the BSM couplings and expected mass values.

\end{abstract}

\maketitle

\thispagestyle{fancy}

\section{Introduction}
The rare B decays are strongly suppressed in Standard Model (SM) due to CKM and by helicity. These decays can be useful for testing the 
New Physics (NP) beyond the SM (BSM). However, the lepton universality observables $R_{K^{(*)}}$ are very useful for testing the NP as the
parameteric uncertainties cancel out at high precision in these ratios. Any small deviation from SM in these measurements will result to 
violation of lepton flavor universality (LFUV), which is a BSM phenomena.\\
Recently LHCb updated the measurement of $R_K$ at Morionod-2019\cite{rk2019} and Belle also presented the result for $R_{K^{*}}$ in $B^{0}$-decays alongwith the counterpart 
$R_{K^{*+}}$ in $B^{+}$-decays\cite{rkstar2019}. These updated results have been included in several global fits\cite{Alguero:2019ptt,Alok:2019ufo,Ciuchini:2019usw,Aebischer:2019mlg}.\\
In this proceedings, we present our results which are reported in detail in ref \cite{Kowalska:2019ley}. We presented the global fit results of Bayesian analysis of the implication of new physics in semileptonic $b \to s $ transitions in model independent approach. We further analyzed a few well-known BSM models and provide the predictions for the BSM couplings and expected mass values.

\section{Fit Methodolgy}
We use the Bayesian approach to constrain the region of NP parameter space which can give a good fit to the data. In this approach, for a theory described by some parameters $m$, experimental observables
$\xi(m)$ can be compared with data $d$ and a pdf $p(m|d)$, of the model parameters $m$, can be calculated through Bayes' Theorem. This reads
\begin{equation}
p(m|d)=\frac{p(d|\xi(m))\pi(m)}{p(d)}\, ,
\label{Bayesth}
\end{equation}
where the likelihood $p(d|\xi(m))\equiv\mathcal{L}(m)$ gives the probability density for obtaining $d$ from a
measurement of $\xi$ given a specific value of $m$, and the prior $\pi(m)$ parametrizes
assumptions about the theory prior to performing the measurement.\\
We define the likelihood function for the set $m$ of input parameters
\begin{equation}
\tiny
 \mathcal{L}(m)= \exp\left\{-\frac{1}{2}\big[\mathcal{O}_{\textrm{th}}(m) -\mathcal{O}_{\textrm{exp}}\big]^T \,
  (\mathcal{C}^{\textrm{exp}} +\mathcal{C}^{\textrm{th}})^{-1} \, \big[\mathcal{O}_{\textrm{th}}(m) -\mathcal{O}_{\textrm{exp}} \big]\right\},
\end{equation}
where $\mathcal{O}_{\rm th}$ and $\mathcal{O}_{\rm exp}$ are theoretical predictions and the experimental measurements observables, respectively. We have taken into account the available experimental correlation which is encoded in the matrix $\mathcal{C}^{\textrm{exp}}$. The $V_{cb}$ element of the CKM matrix is treated as a real 
nuisance parameter. We scan it together with the models' input parameters, following a Gaussian distribution 
around its central Particle Data Group (PDG) value, and adopting PDG uncertainties. We always scan NP wilson coefficient from $-3$ to $3$.

\section{Results}
The effective Hamiltonian for the $b \to s ll$ transition can be written as:
\be \label{heff}
\mathcal{H}_{eff}=-\frac{4G_F}{\sqrt{2}}V_{tb}V_{ts}^* \frac{e^2}{16\pi^2}\sum_{i,l}(C_i^l O_i^l + C_i^{' l} O_i^{' l}) + \textrm{H.c.}\,,
\ee
In this study, we assume the presence of NP in the following semi-leptonic operators:
\bea
O_9^{(\prime)l}&=&(\bar{s}_{L(R)}\gamma^\mu b_{L(R)})(\bar{l}\gamma_\mu l)\\
O_{10}^{(\prime)l}&=&(\bar{s}_{L(R)}\gamma^\mu b_{L(R)})(\bar{l}\gamma_\mu\gamma_5 l)
\eea
where the lepton $l$ can be an electron or a muon. The full list of observables included in the fit can be found in ref. \cite{Kowalska:2019ley}. 

\subsection{Model Independent Analysis}

\begin{figure}[h!]
  \centering
\mbox{
  \subfigure{\includegraphics[scale=0.2]{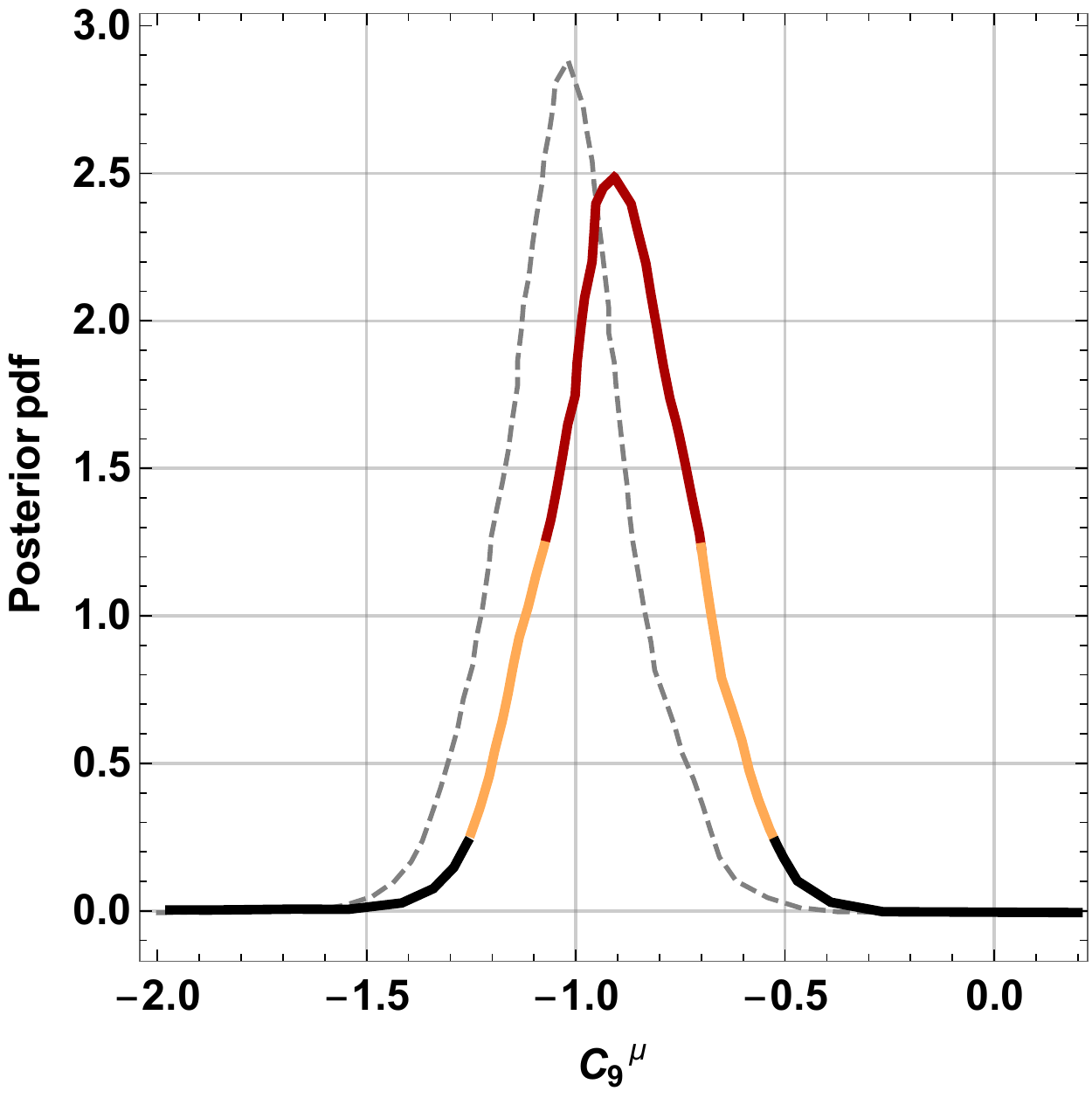}}\quad
  \subfigure{\includegraphics[scale=0.2]{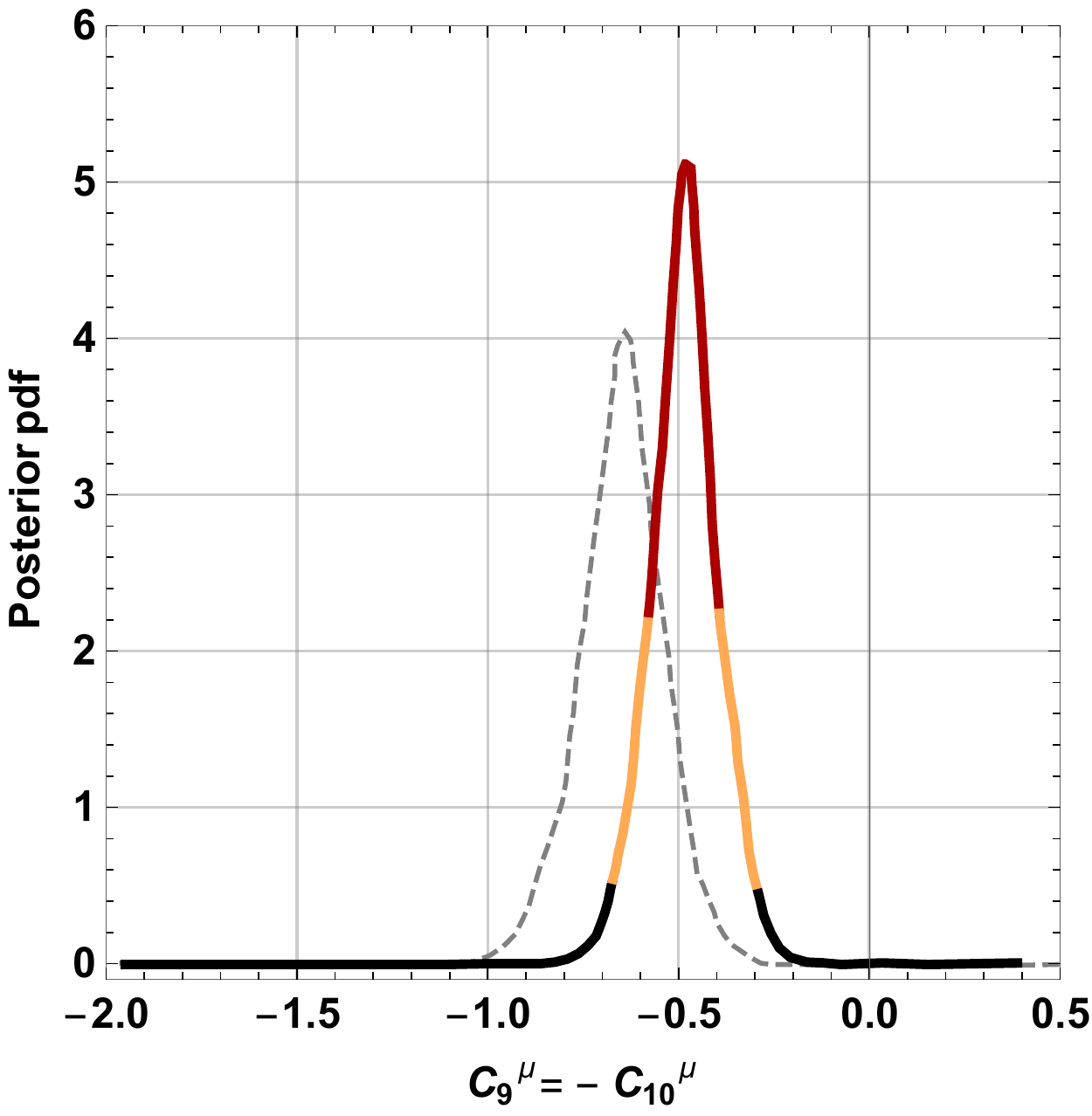}}
}
\caption{\footnotesize (a) Posterior pdf for $C_9^{\mu}$(left) and $C_9^{\mu}=-C_{10}^{\mu}$ (right).
}
\label{fig:1par}
\end{figure}

\begin{figure}[h!]
\centering
\mbox{
\subfigure{\includegraphics[scale=0.1]{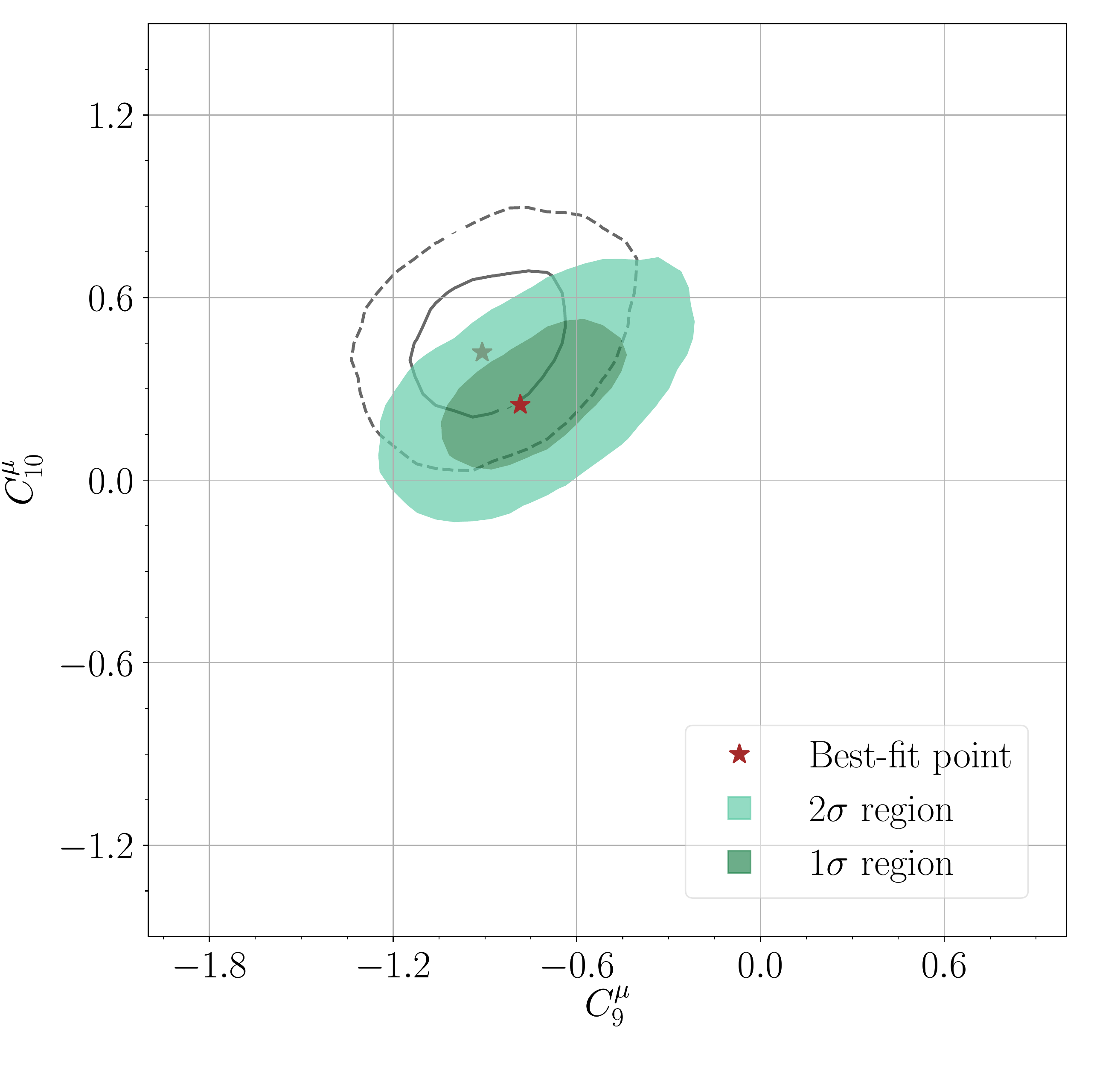}}\quad
\subfigure{\includegraphics[scale=0.1]{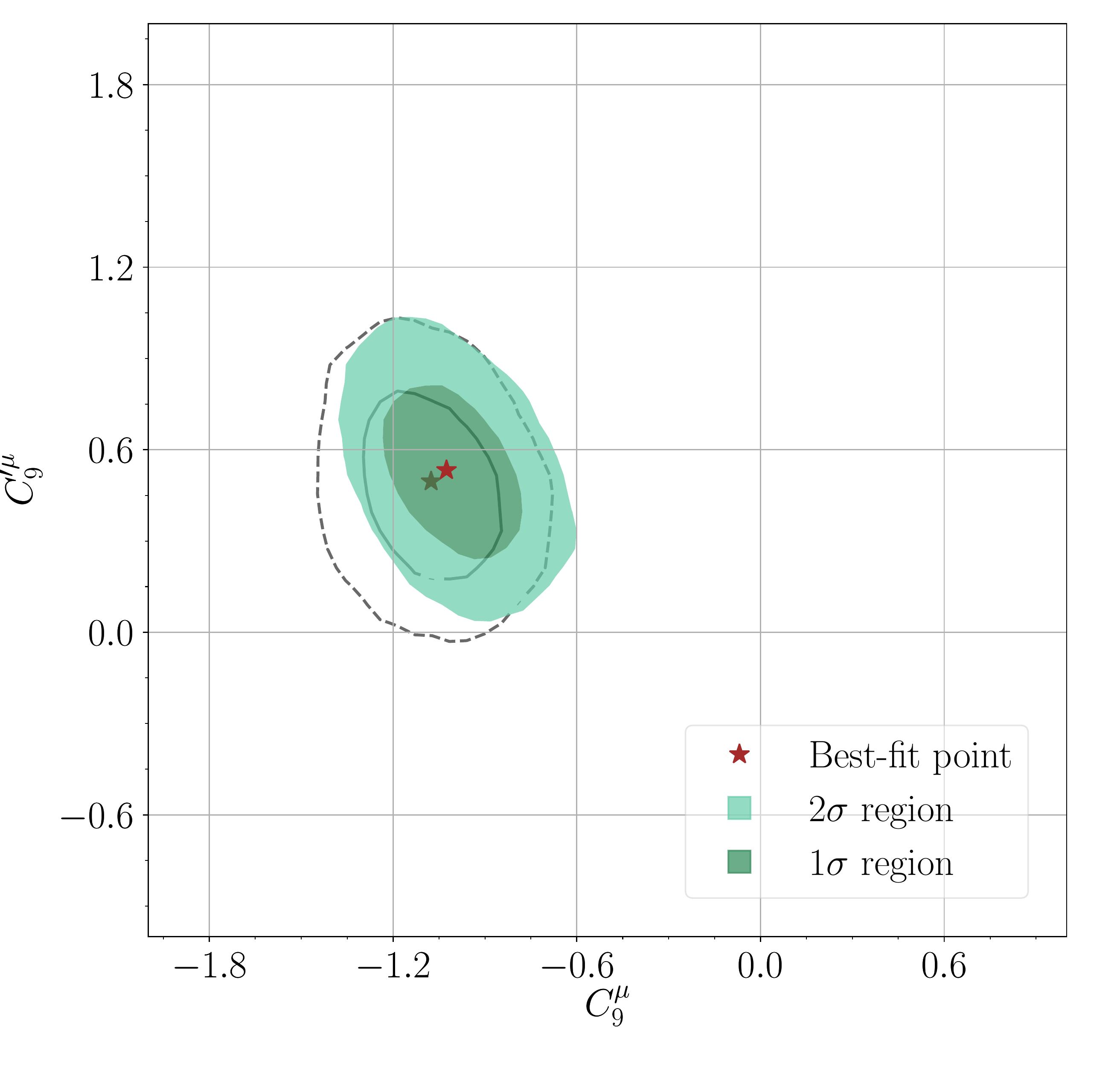}}
}
\caption{\footnotesize Posterior pdf for $(C_9^{\mu}$, $C_{10}^{\mu})$ (left) and $(C_9^{\mu}$, $C_{9}^{\prime\mu})$ (right).
\label{fig:2pars}}
\end{figure}

\begin{figure}[h!]
\centering
\mbox{
\subfigure{\includegraphics[scale=0.09]{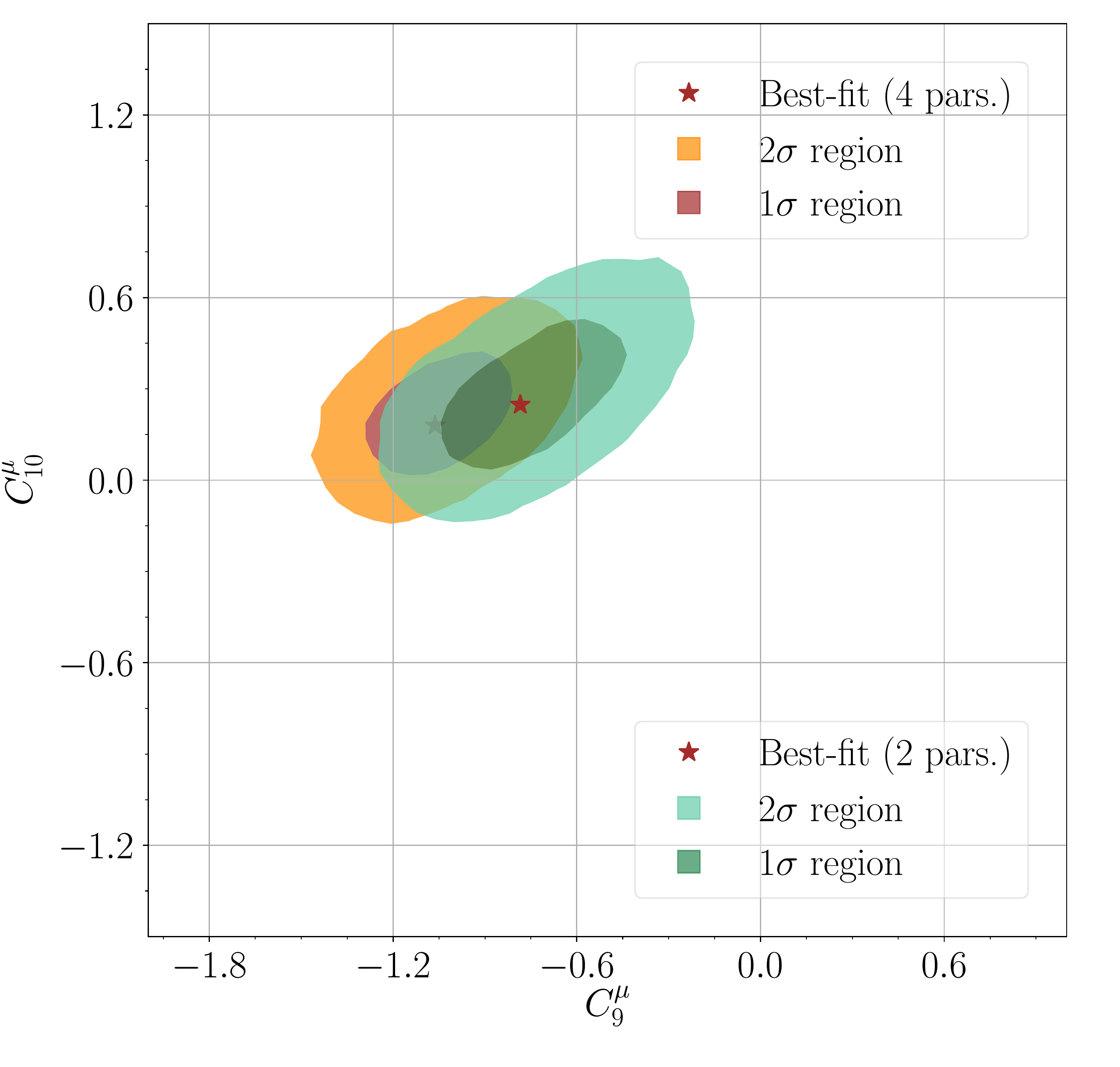}}\quad
\subfigure{\includegraphics[scale=0.09]{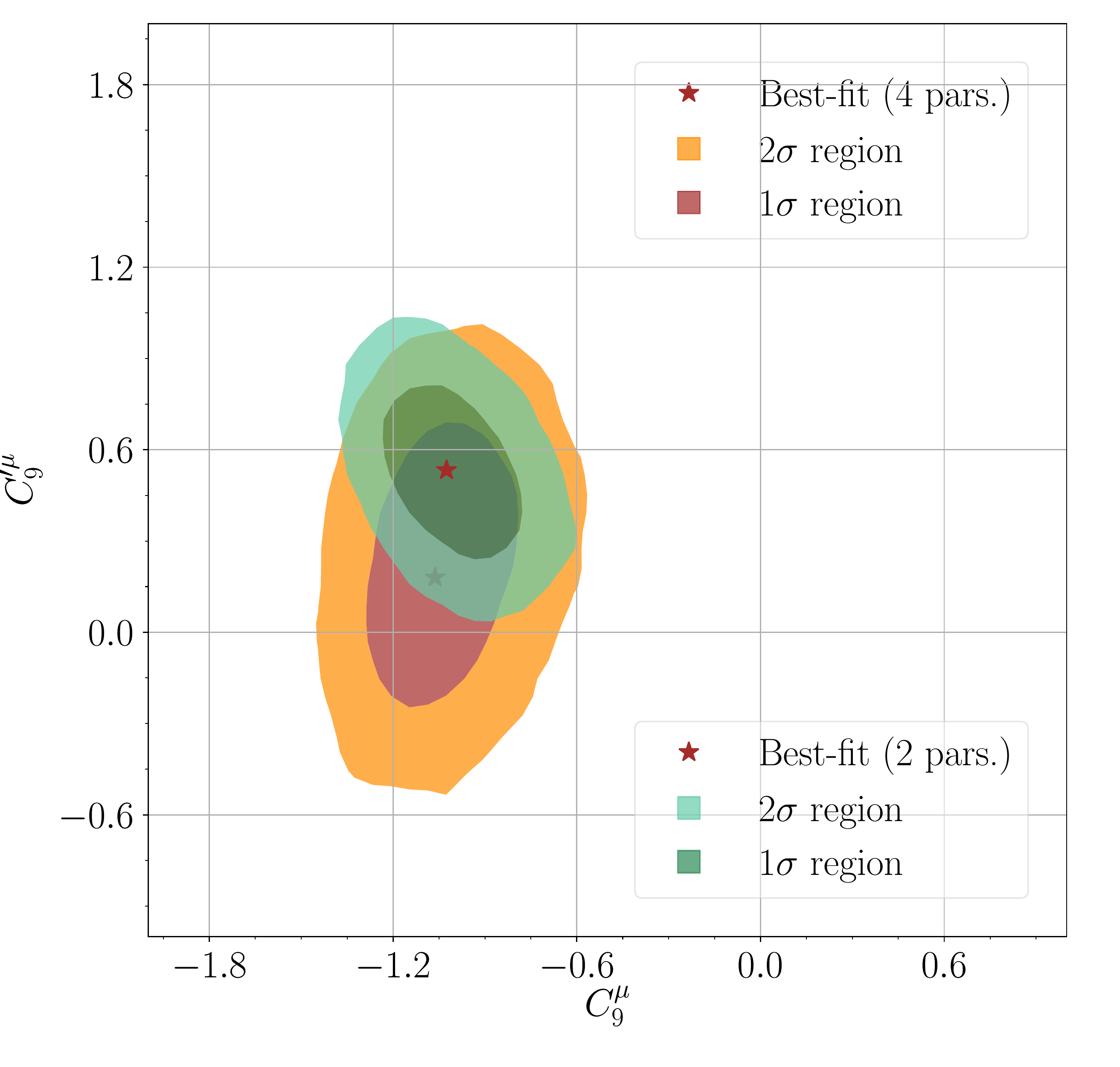}}\quad
\subfigure{\includegraphics[scale=0.125]{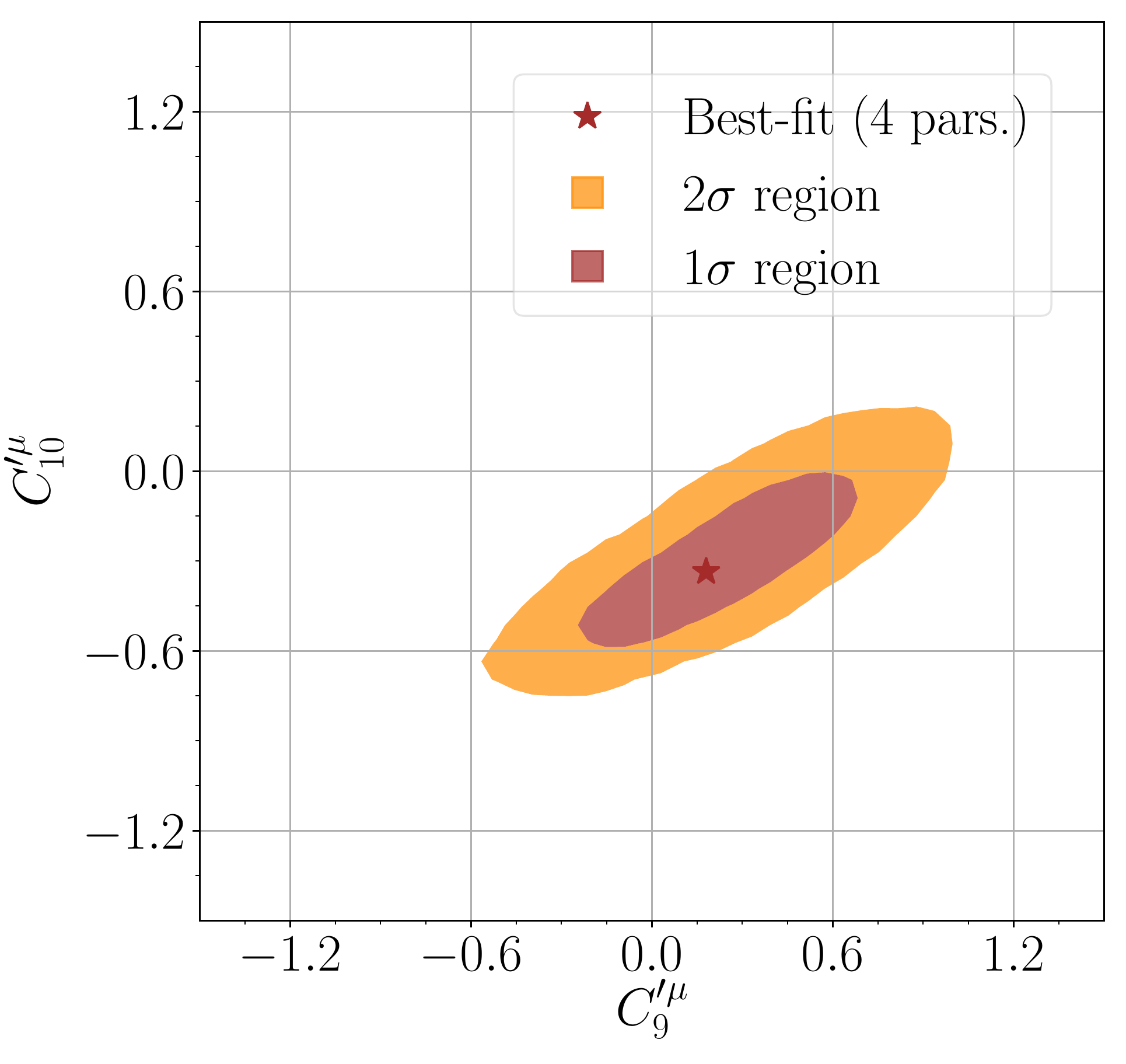}}
}
\caption{\footnotesize Comparison of posterior pdf in 2 parameters and 4 parameters scan. The explicit correlation for $C_9^{\prime\mu}$ and $C_{10}^{\prime\mu}$ is also shown (with 4 parameter scan)
\label{fig:4pars}}
\end{figure}

\begin{figure}[t]
\centering
\mbox{%
\subfigure{\includegraphics[scale=0.1]{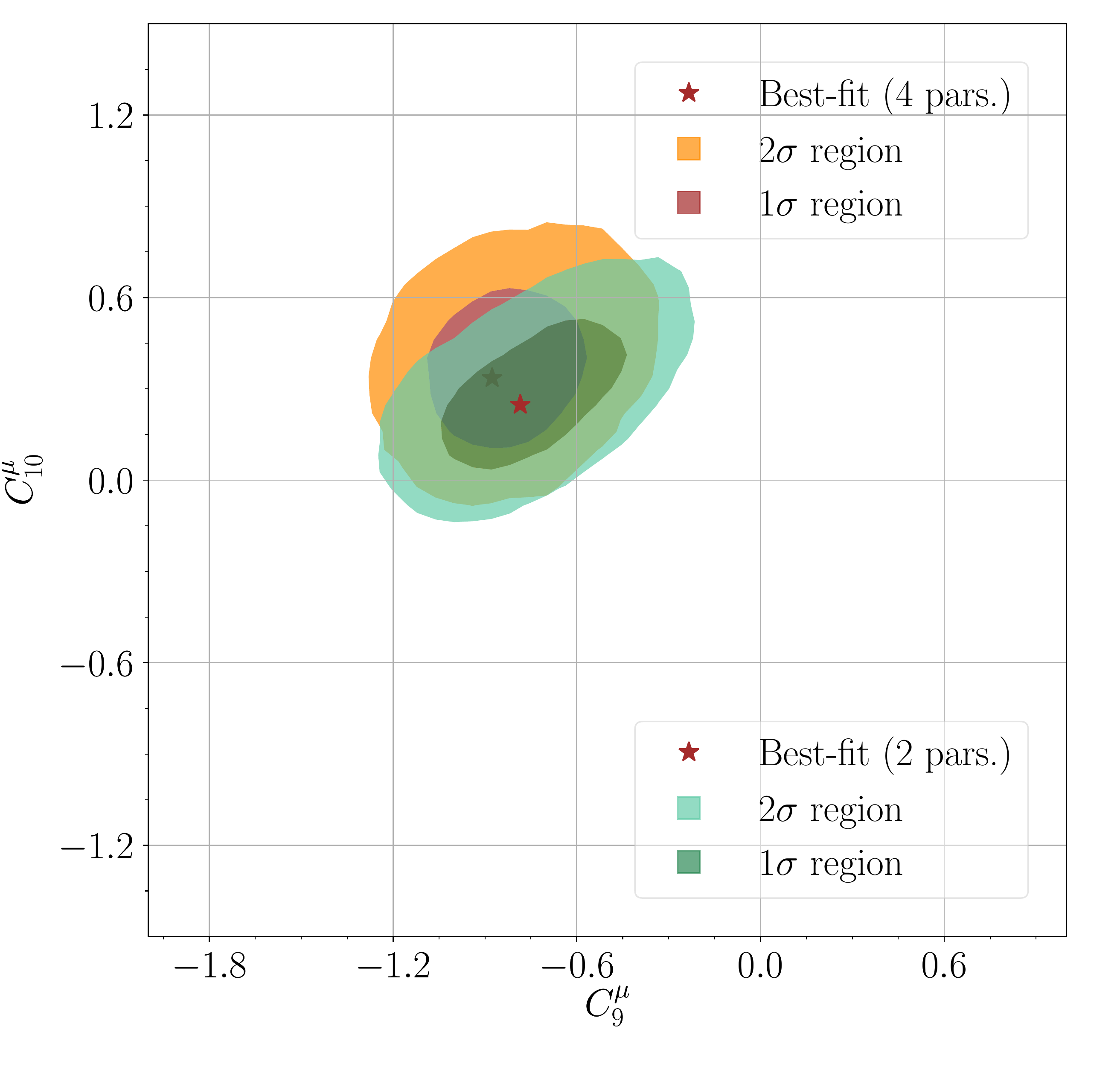}}\quad
\subfigure{\includegraphics[scale=0.1]{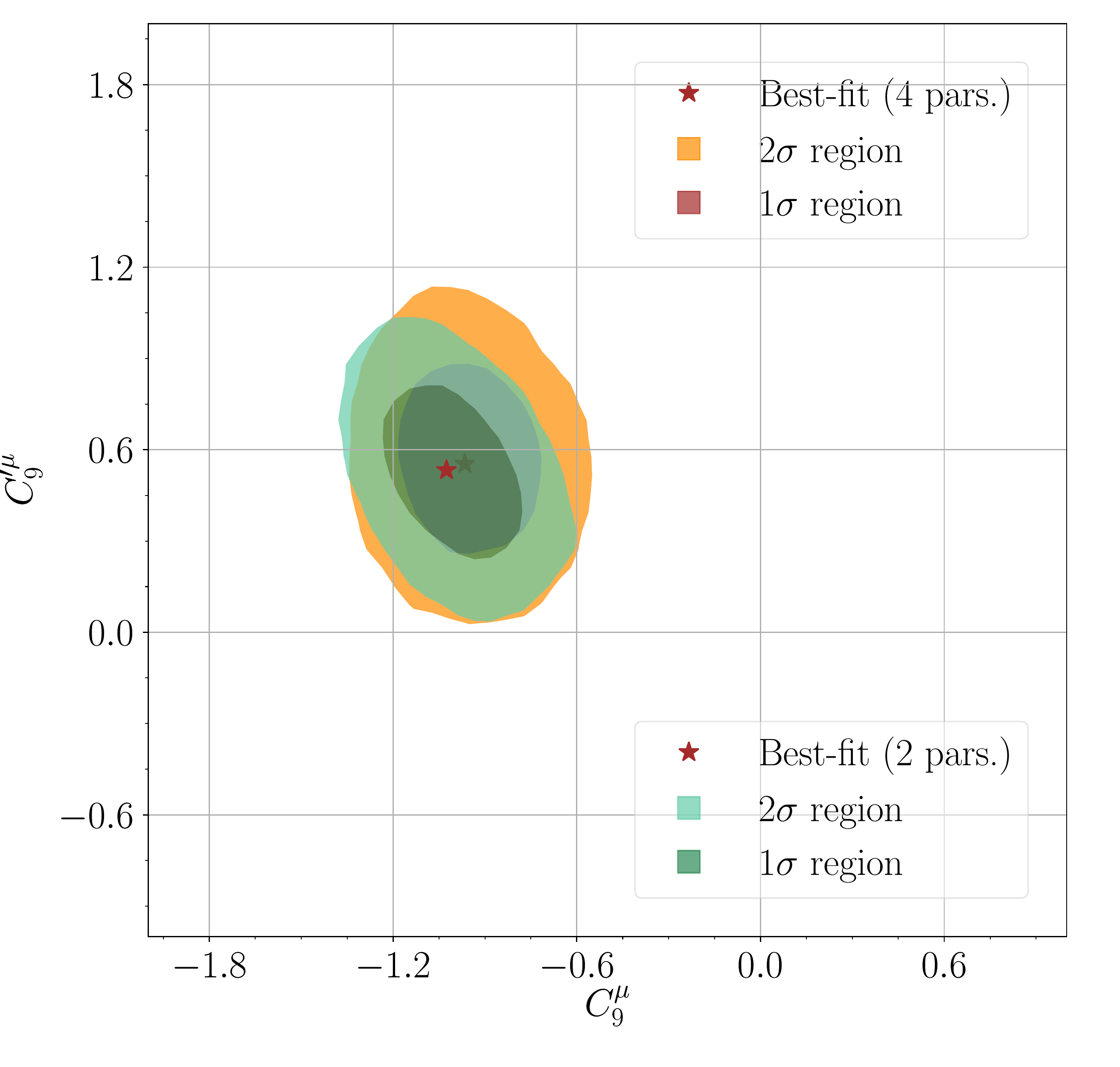}}
}%
\\
\mbox{%
\subfigure{\includegraphics[scale=0.14]{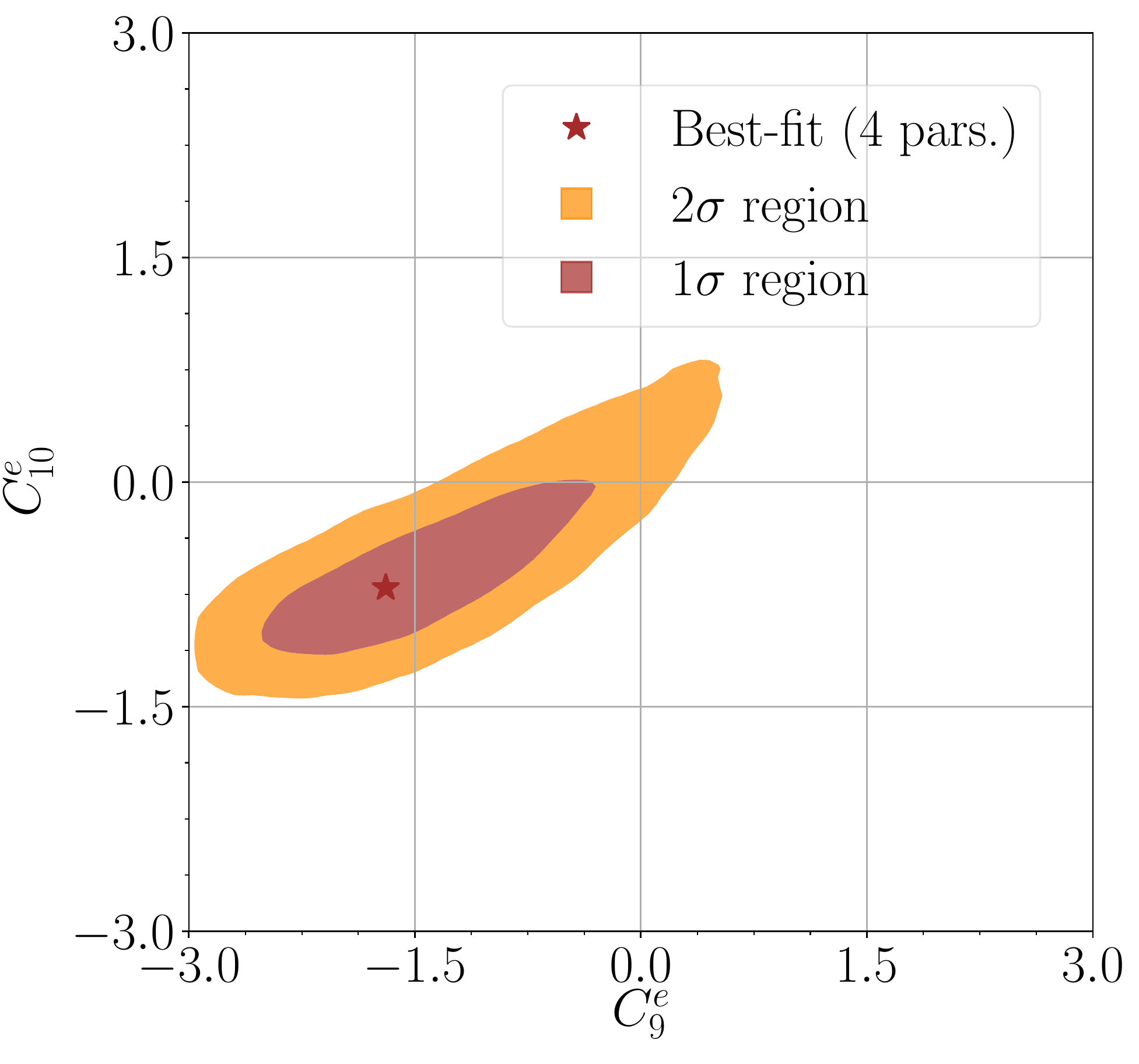}}\quad
\subfigure{\includegraphics[scale=0.14]{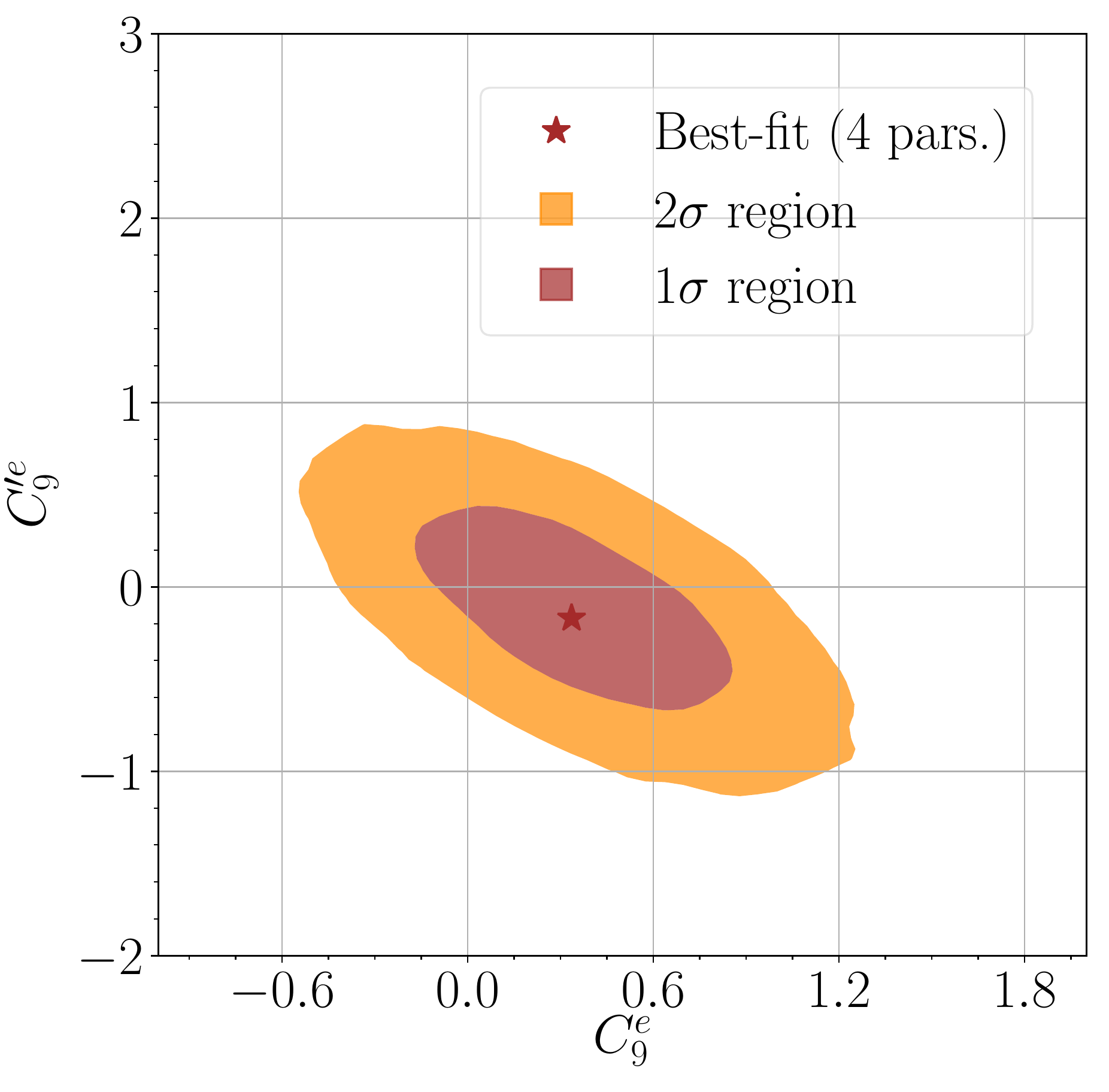}}
}%
\caption{\footnotesize Comparison of posterior pdf in 2 parameters and 4 parameters scan with NP in electron sector. The explicit correlation for only electron sector WC's in 4 paraemeters scan. 
\label{fig:4parsb}}
\end{figure}

\begin{figure}[t]
\centering
\mbox{%
\subfigure{\includegraphics[scale=0.1]{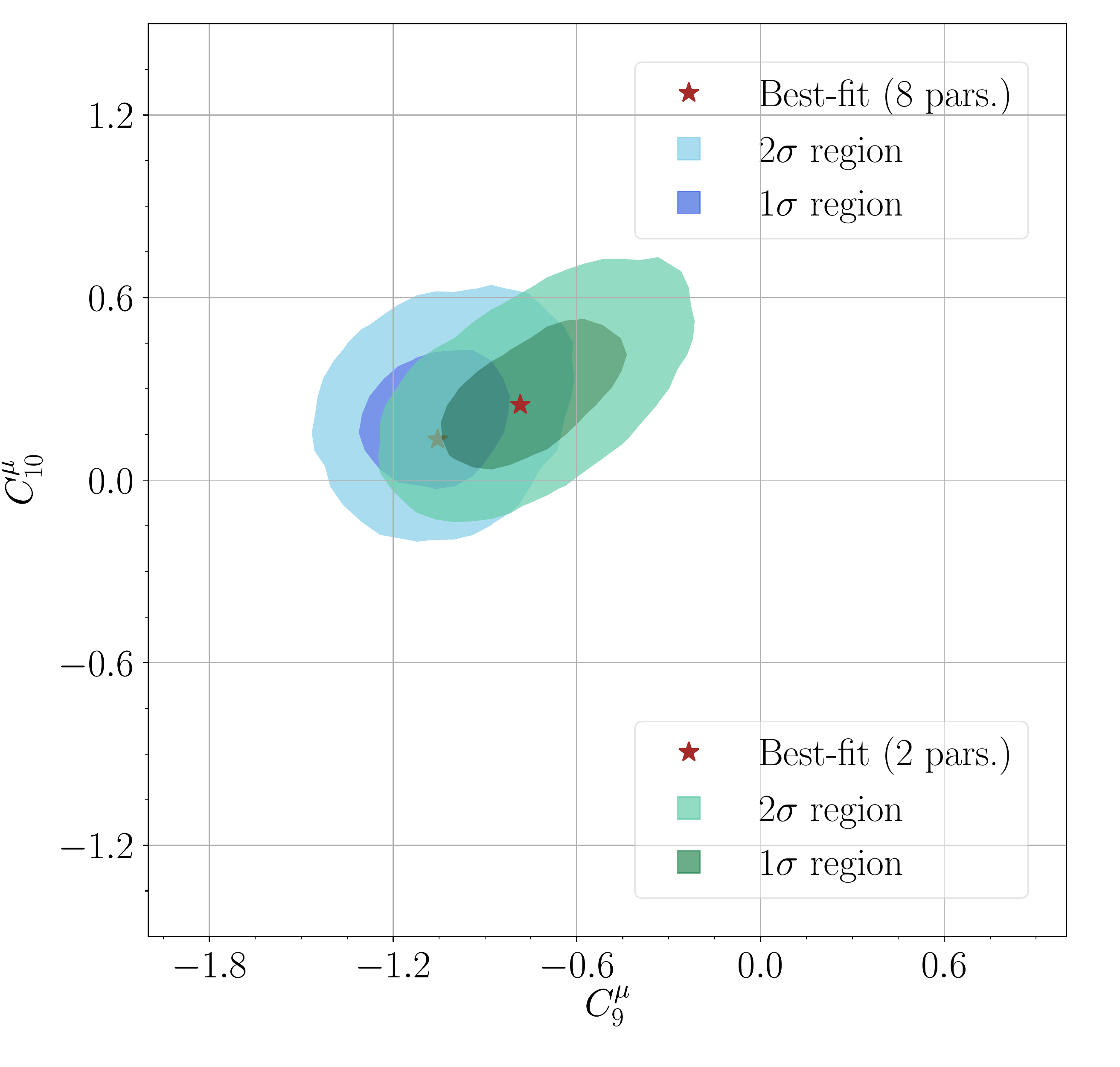}}\quad
\subfigure{\includegraphics[scale=0.1]{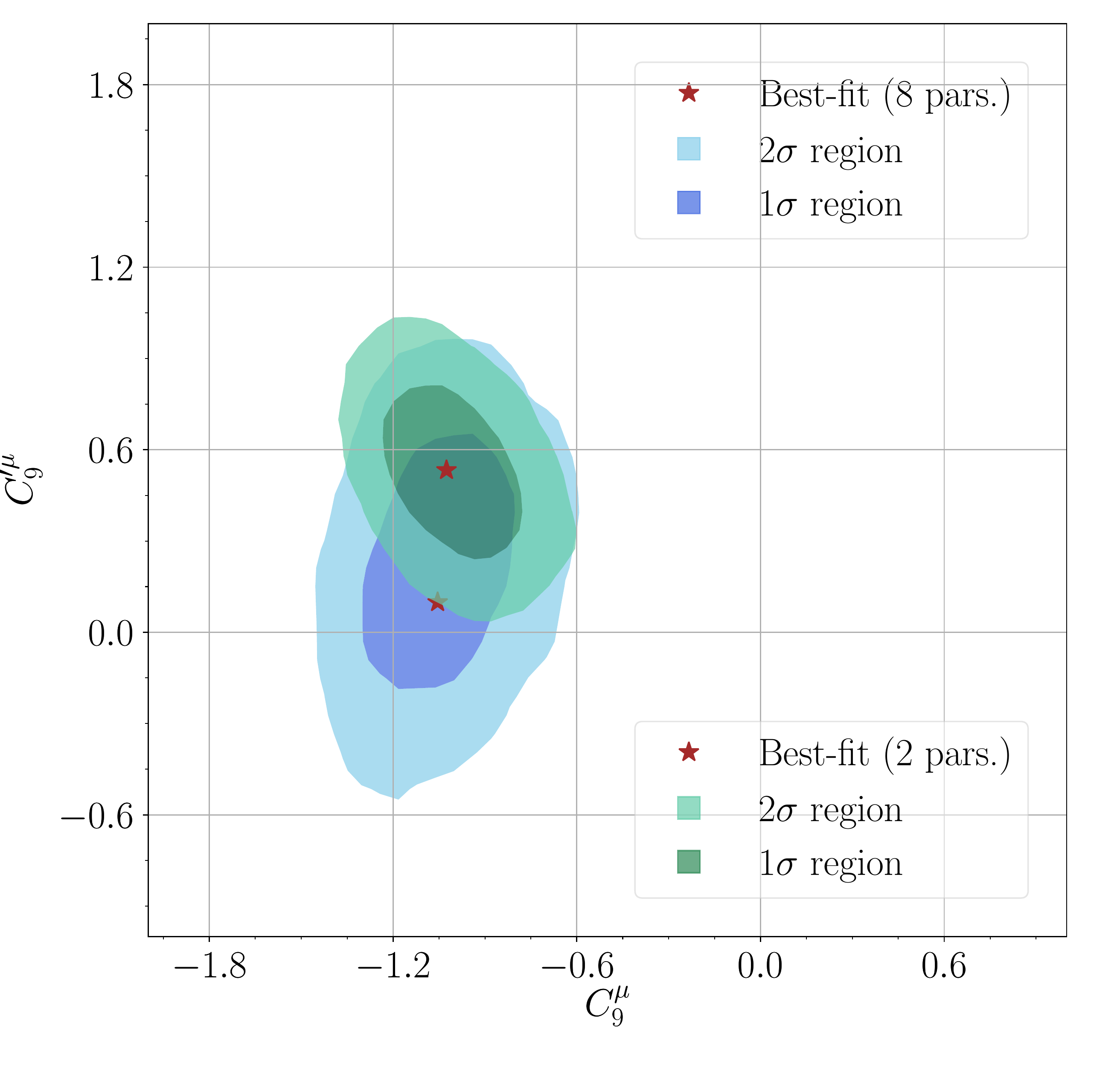}}
}%
\caption{\footnotesize Comparison of posterior pdf in 2 parameters versus 8 parameters scan.
\label{fig:8pars}}
\end{figure}

We present the posterior pdf of single non-zero NP wilson coefficient $C_9^{\mu}$ (left panel of figure \ref{fig:1par}) and $C_9^{\mu}=-C_{10}^{\mu}$ (right panel of figure \ref{fig:1par}) marginalized over the nuisance parameter.The red and orange color represent the $1\,\sigma$ and $2\,\sigma$ credible regions, respectively. The gray dashed line shows the posterior pdf corresponding to the data pre-LHCb Run~2.\\
In the left panel of Fig.\ref{fig:2pars}, the posterior pdf for the scan in the input parameter $(C_9^{\mu}$, $C_{10}^{\mu})$ is presented. The red star marks the position of the best-fit point. The gray solid (dashed) line shows the $1\,\sigma$ ($2\,\sigma$) credible region of the pdf corresponding to the data pre-LHCb Run~2. 
The associated best-fit point is also shown in gray. The new measurement of $R_K$, which is slightly higher than the previous measurement, brings the $2\,\sigma$ region closer to the axes origin. In this case, in fact, one expects $R_K \approx R_{K^*}$ and a tension between the measurements of $R_K$ and $R_{K^*}$ arises as the posterior pdf becomes narrower. In the right panel of Fig.\ref{fig:2pars}, the posterior pdf for the scan in the input parameter $C_9^{\mu}$, $C_{9}^{\prime\mu}$ is presented.\\
We performed a scan with 4 NP parameters $C_9^{\mu}$, $C_{10}^{\mu}$, $C_9^{\prime\mu}$, $C_{10}^{\prime\mu}$ and make the comparison between the marginalized pdf in the 
($C_9^{\mu}$, $C_{10}^{\mu}$) plane for the  scan with 2 input NP parameters, 
and the one with 4 NP parameters which is shown in the left of Figure \ref{fig:4pars}. The large negative values of $C_9^{\mu}$ are favored by the data with 4 parameters. In the middle of Figure \ref{fig:4pars}, we show a comaprison between the marginalized pdf in the 
($C_9^{\mu}$, $C_{9}^{\prime\mu}$). It can be seen that ample region of $C_9^{\prime\mu}\leq 0$ is allowed due to the introduction of $C_{10}^{\prime\mu}$. The explicit correlation between the $C_9^{\prime\mu}$ and $C_{10}^{\prime\mu}$ in the right of Figure \ref{fig:4pars} in case of scan with 4 parameters.
\\
In upper left panel of Figure \ref{fig:4parsb}, we show the $1\,\sigma$ (dark) and $2\,\sigma$ (light) credible regions of the posterior pdf for the scan in the input parameter $C_9^{\mu}$, $C_{10}^{\mu}$, compared with the marginalized 2-dimensional regions in the same parameters for 
the scan with $C_9^{\mu}$, $C_{10}^{\mu}$, $C_9^{e}$, $C_{10}^{e}$ all floating, 
which are shown in brown ($1\,\sigma$) and orange ($2\,\sigma$). A similar comparison of the posterior pdf for the scan in $C_9^{\mu}$, $C_9^{\prime\mu}$ and the one 
with $C_9^{\mu}$, $C_{9}^{\prime\mu}$, $C_9^{e}$, $C_{9}^{\prime e}$ all floating in the upper right panel of Figure \ref{fig:4parsb}. In the lower panel of Figure \ref{fig:4parsb}, a marginalized pdf for electron sector Wilson coefficients are presented which is consistent with zero at $2\sigma$. This suggests that NP with only muon sector can easily explain the present data. 
\\
In Figure \ref{fig:8pars}, we present the marginalized pdf for 8 parameter scan in most relevant planes $(C_9^{\mu}$, $C_{10}^{\mu})$ and $(C_9^{\mu}$, $C_9^{\prime\mu})$ in the left and right panel. These marginal pdf are compared with 2 parameter scan and we find that these figures are almost same as Figure \ref{fig:4pars} which is expected as the NP wilson coefficients in the electron sector have limited impact on the data.

\begin{table}[t]
	\begin{center}\tiny
		\begin{tabular}{|c|c|ccc|cccc|}
         \toprule                     
           Input parameters & $\textbf{ln}$ $\mathcal{Z}$ & $\bf Pull$ & $\chi^2_{\textrm{TOT}}$ & $\frac{\chi^2_{\textrm{TOT}}}{d.o.f}$& $\chi^2_{\mu}$ &$\chi^2_{e}$ & $\chi^2_{R_{K}}$ & $\chi^2_{R_{K^*}}$ \\
		\toprule
            \bf SM  &$88.5$ & $-$ &  $174.7$ & $1.29$& $145.7$ & $6.5$ & $8.1$ & $12.0$  \\
            & $88.3$ & $-$ & $174.4$ & $1.24$&  $145.7$ & $6.5$ & $6.2$ & $13.6$ \\
            \hline
            $C_9^{\mu}$ &  $75.8$  & $5.0\,\sigma$ &$145.6$ &$1.09$ & $132.5$ & $6.7$ & $0.2$ & $6.0$\\
            & $77.3$ &$4.7\,\sigma$ & $148.4$ & $1.06$&  $132.2$ & $6.6$ & $0.3$ & $8.9$ \\ 
            \hline
            $C_9^{\mu}=-C_{10}^{\mu}$ &  $74.4$  & $5.3\,\sigma$ &$142.4$ &$1.06$ & $132.4$ & $6.8$ & $0.2$ & $3.0$\\
            
            & $77.5$ &$4.8\,\sigma$ & $148.2$ & $1.06$&  $133.2$ & $6.7$ & $1.2$ & $7.0$ \\            
            \hline
            $C_9^{\mu},\,C_{10}^{\mu}$ & $74.5$& $5.3\,\sigma$ & $140.1$ & $1.05$ & $129.8$ & $6.8$ & $0.2$& $3.4$  \\
             & $77.6$ &$4.7\,\sigma$ & $146.1$ & $1.05$ &  $130.3$ & $6.7$ & $1.5$ & $7.6$\\
            \hline
            $C_9^{\mu},\,C_{9}^{\prime\mu}$ & $75.1$ & $5.2\,\sigma$ & $141.1$ & $1.06$& $128.1$ & $6.7$ & $2.0$& $4.1$ \\
             &$75.8$ &$5.0\,\sigma$& $142.3$ & $1.02$& $127.6$ & $6.7$ & $0.5$ & $7.3$  \\
            \hline
            $C_{9}^{\mu},\,C_{10}^{\mu},\,C_{9}^{\prime\mu},\,C_{10}^{\prime\mu}$ & $74.0$ & $5.4\,\sigma$ &$133.3$ & $1.02$  &$123.5$ & $6.8$& $0.6$ & $2.4$\\
             &$76.0$ & $5.1\,\sigma$ & $136.8$ & $1.00$&$123.2$ & $6.8$ & $0.0$ & $6.8$ \\
            \hline
            $C_{9}^{\mu},\,C_{10}^{\mu},\,C_{9}^{e},\,C_{10}^{e}$ & $75.6$ &  $4.9\,\sigma$ & $138.8$ & $1.06$& $129.7$ & $6.9$ & $0.0$ & $2.1$ \\
           
            & $78.0$& $4.5\,\sigma$&$142.7$ & $1.04$ &$129.8$ & $7.1$& $0.1$ & $5.8$ \\
            \hline
            $C_{9}^{\mu},\,C_9^e,\,C_{9}^{\prime \mu},\,C_{9}^{\prime e}$ & $75.8$ & $4.9\,\sigma$ & $138.5$ & $1.06$ & $127.5$ & $7.8$ & $0.5$ & $2.4$\\
           
             & $77.7$& $4.6\,\sigma$ & $141.6$ &$1.03$ & $127.2$ & $7.0$& $0.2$& $6.7$\\
            \hline
            $(C_{9}^{\mu},\,C_{10}^{\mu},\,C_{9}^{\prime\mu},\,C_{10}^{\prime\mu}$ & $76.2$ &$4.7\,\sigma$ &$132.4$ &$1.04$ & $123.3$ & $6.7$ & $0.3$ & $2.1$\\
            $C_{9}^{e},\,C_{10}^{e},\,C_{9}^{\prime e},\,C_{10}^{\prime e})$ & $78.3$ &$4.4\,\sigma$ &$135.4$ & $1.02$& $123.3$& $6.6$& $0.2$ & $ 5.4$\\
            \toprule
		\end{tabular}
		\caption{Evidence, pull from the SM, and chi-squared statistics for the best-fit points of the considered scenarios. Second row in each block correspond to the new data, while the first ones show the previous determinations.}
		\label{tab:results_chi2}
	\end{center}
\end{table}


\begin{table}[t]
\footnotesize
\centering
\tiny{
		\begin{tabular}{|c|cccccccc|}
		    \toprule
            \bf{Input parameters} & $ C_9^{\mu}$ & $ C_{10}^{\mu}$ & $ C_9^{\prime\mu}$ & $ C_{10}^{\prime\mu}$ & $ C_9^{e}$ & $ C_{10}^{e}$ & $ C_9^{\prime e}$ & $ C_{10}^{\prime e}$ \\
		    \toprule
		    ${C_9^{\mu}}$ & $-1.02$ & $0$ & $0$ & $0$ & $0$ & $0$  & $0$  & $0$\\
             & $-0.90$ & $0$ & $0$ & $0$ & $0$ & $0$  & $0$  & $0$  \\
            \hline
		    $ C_9^{\mu}=-C_{10}^{\mu}$ & $-0.64$ & $0.64$ & $0$ & $0$ & $0$ & $0$  & $0$  & $0$ \\            
            & $-0.48$ & $0.48$ & $0$ & $0$ & $0$ & $0$  & $0$  & $0$  \\
\hline
            $C_9^{\mu},\,C_{10}^{\mu}$ & $-0.91$ & $0.42$ & $0$ & $0$ & $0$ & $0$  & $0$  & $0$ \\
           
             & $-0.78$ & $0.25$ & $0$ & $0$ & $0$ & $0$  & $0$  & $0$ \\
            \hline
            $C_9^{\mu},\,C_{9}^{\prime\mu}$ & $-1.08$ & $0$ & $0.49$ & $0$ & $0$ & $0$  & $0$  & $0$\\
            
            & $-1.03$ & $0$& $0.53$& $0$& $0$& $0$&$0$ &$0$ \\
            \hline
            $C_{9}^{\mu},\,C_{10}^{\mu},\,C_{9}^{\prime\mu},\,C_{10}^{\prime\mu}$ & $-1.14$& $0.28$ &$0.21$ &$-0.31$ & $0$& $0$& $0$& $0$ \\
       
             & $-1.06$ & $0.18$ & $0.18$ &$-0.34$ & $0$& $0$& $0$& $0$\\
            \hline
            $C_{9}^{\mu},\,C_{10}^{\mu},\,C_{9}^{e},\,C_{10}^{e}$ & $-0.92$ & $0.40$ & $0$ & $0$ & $-1.50$& $-0.90$ & $0$ & $0$ \\
             
            & $-0.88$& $0.34$& $0$& $0$& $-1.69$& $-0.71$& $0$&$0$ \\
            \hline
            $C_{9}^{\mu},\,C_9^e,\,C_{9}^{\prime \mu},\,C_{9}^{\prime e}$ & $-1.02$ & $0$ & $0.54$ & $0$ & $0.58$& $0$ & $-0.17$ & $0$ \\
             
            &$-0.97$ & $0$&$0.55$ &$0$ & $0.34$&$0$ & $-0.17$& $0$\\
            \hline
            $(C_{9}^{\mu},\,C_{10}^{\mu},\,C_{9}^{\prime\mu},\,C_{10}^{\prime\mu}$ &$-1.10$ & $0.21$&  $0.21$ & $-0.30$& $-0.80$& $-0.63$ & $-0.73$& $-0.57$ \\
      
            $C_{9}^{e},\,C_{10}^{e},\,C_{9}^{\prime e},\,C_{10}^{\prime e})$ & $-1.05$&$0.13$& $0.10$ &$-0.38$& $-2.18$ & $-0.07$&$-2.73$& $-1.34$ \\
           
            \hline
		\end{tabular}}
		\caption{Wilson coefficients at the best-fit points, as well as the values there of $R_K$ and $R_{K^{\ast}}$. Second row in each correspond to the new data, while the first ones show the previous determinations.}
		\label{tab:results_BF} 
\end{table}
We use Jeffrey's scale to quickly assess the Bayes factor, which will point to which model is favored by the data. We find that models with scenario $(C_9^{\mu},C_9^{\prime\mu})$ and $C_{9}^{\mu},\,C_{10}^{\mu},\,C_{9}^{\prime\mu},\,C_{10}^{\prime\mu}$ are slightly favored by the data. We have summarized all 8 scans in Table \ref{tab:results_chi2}. In order to make contact with frequentist approach, the best fit values  of wilson coefficeints with $R_K$ and $R_{K^{\ast}}$ at best fit points is presented in Table \ref{tab:results_BF}.

\section{Model dependent analysis}

\subsection{Heavy $Z^{\prime}$}
The most generic Lagrangian, parametrizing LFUV couplings of $Z'$ to the $b$-$s$ current and the muons reads
\bea
\mathcal{L}\supset  && Z'_{\alpha}\left(\Delta_L^{sb}\, \bar{s}_L\gamma^{\alpha}\,b_L+ \Delta_R^{sb}\, \bar{s}_R\gamma^{\alpha}\,b_R+\textrm{H.c.}\right)\,\nonumber\\ 
&&+Z'_{\alpha}\left(\Delta^{\mu\mu}_{L}\,\bar{\mu}_L\gamma^{\alpha}\mu_L+\Delta^{\mu\mu}_{R}\,\bar{\mu}_R\gamma^{\alpha}\mu_R\right)\,.
\eea
The relevant Wilson coefficients are then given by
\be\label{C9generic}
\tiny
C_{9,\textrm{NP}}^{(\prime)\mu}=-2\frac{\Delta_{L(R)}^{sb}\Delta^{\mu\mu}_9}{V_{tb}V_{ts}^{\ast}}\left(\frac{\Lambda_v}{m_{Z'}}\right)^2, 
C_{10,\textrm{NP}}^{(\prime)\mu}=-2\frac{\Delta_{L(R)}^{sb}\Delta^{\mu\mu}_{10}}{V_{tb}V_{ts}^{\ast}}\left(\frac{\Lambda_v}{m_{Z'}}\right)^2\,,
\ee
where $\Delta_9^{\mu\mu}\equiv (\Delta^{\mu\mu}_{R}+\Delta^{\mu\mu}_{L})/2$, $\Delta_{10}^{\mu\mu}\equiv (\Delta^{\mu\mu}_{R}-\Delta^{\mu\mu}_{L})/2$, $m_{Z'}$ 
is the mass of the $Z'$ boson, and
$\Lambda_v=\left(\frac{\pi}{\sqrt{2}G_F\alpha_{\textrm{em}}}\right)^{1/2}\approx 4.94\tev$,
is the typical effective scale of the new physics.

The coupling of heavy $Z'$ to the gauge eigenstates must be flavor-conserving if it is the gauge boson of a new U(1)$_X$ gauge group and 
an additional structure is required to generate $\Delta_{L}^{sb}$ and $\Delta_{R}^{sb}$. Thus, in this work we also 
consider the impact of the new LHCb and Belle data 
on the masses and couplings of a few simplified but UV complete models.

\textbf{Model~1.} We consider a U(1)$_X$ model that has proven to be quite popular is the 
traditional $X=L_{\mu}-L_{\tau}$ model. Besides $Z'$, we also add to the SM a scalar singlet field $S$ to spontaneously break the U(1)$_X$ symmetry and 
VL quark pairs $Q,Q^{\prime}$ and $D,D^{\prime}$ to create the flavor-changing couplings $\Delta^{bs}_{L,R}$\cite{Fox:2011qd,Bobeth:2016llm}.

\textbf{Model~2.} Another realization of the $L_{\mu}-L_{\tau}$ model we consider is 
an extension of the SM characterized by one pair of VL quark doublets $Q,Q^{\prime}$, 
to generate the flavor-violating coupling of the $Z'$ 
in the quark sector, $\Delta^{bs}_L$, and one pair of VL U(1)$_X$ neutral leptons $E,E'$, 
which have to be SU(2) singlets\cite{Altmannshofer:2016oaq,Darme:2018hqg}.

\textbf{Model~3.} We finally consider an alternative to the $L_{\mu}-L_{\tau}$ model, obtained if one 
charges the VL leptons under the U(1)$_X$ symmetry, and leaves the SM leptons uncharged\cite{Sierra:2015fma}.\\

The gauge quantum numbers of the additional fermions and the contribution to the NP wilson coefficients in these models can be read from ref.\cite{Kowalska:2019ley}\\
\begin{figure}[h]
\centering
\mbox{%
\subfigure{\includegraphics[scale=0.13]{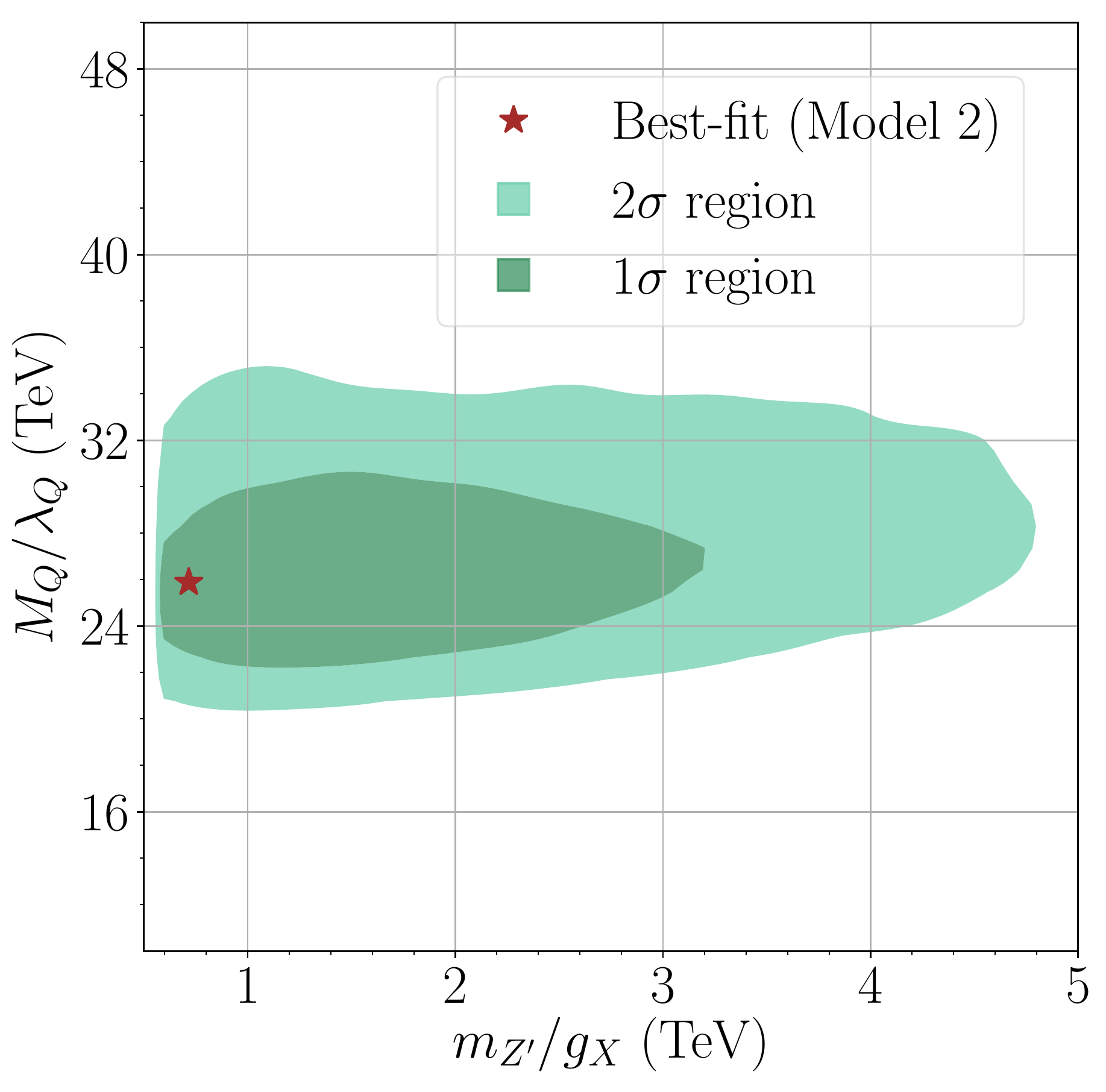}}\quad
\subfigure{\includegraphics[scale=0.345]{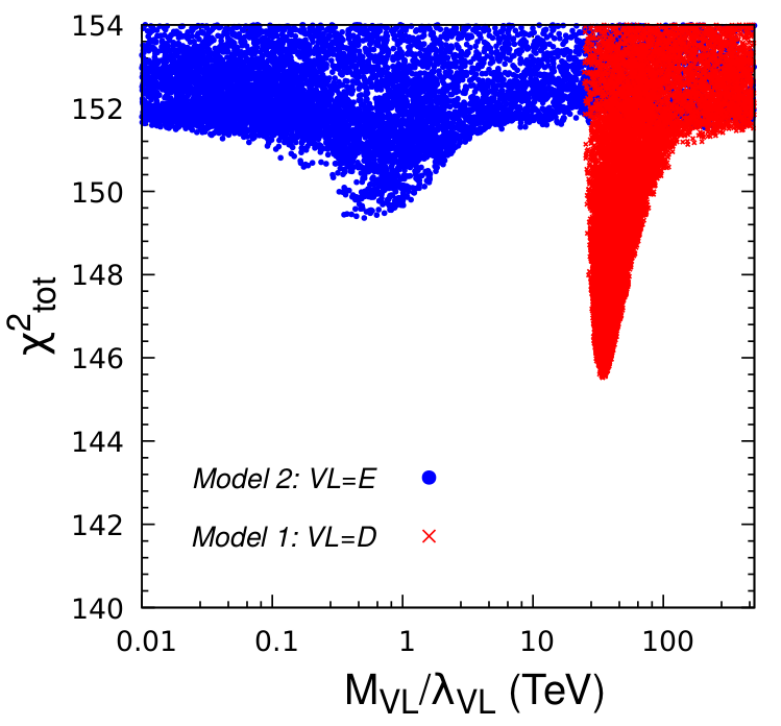}}\quad
\subfigure{\includegraphics[scale=0.13]{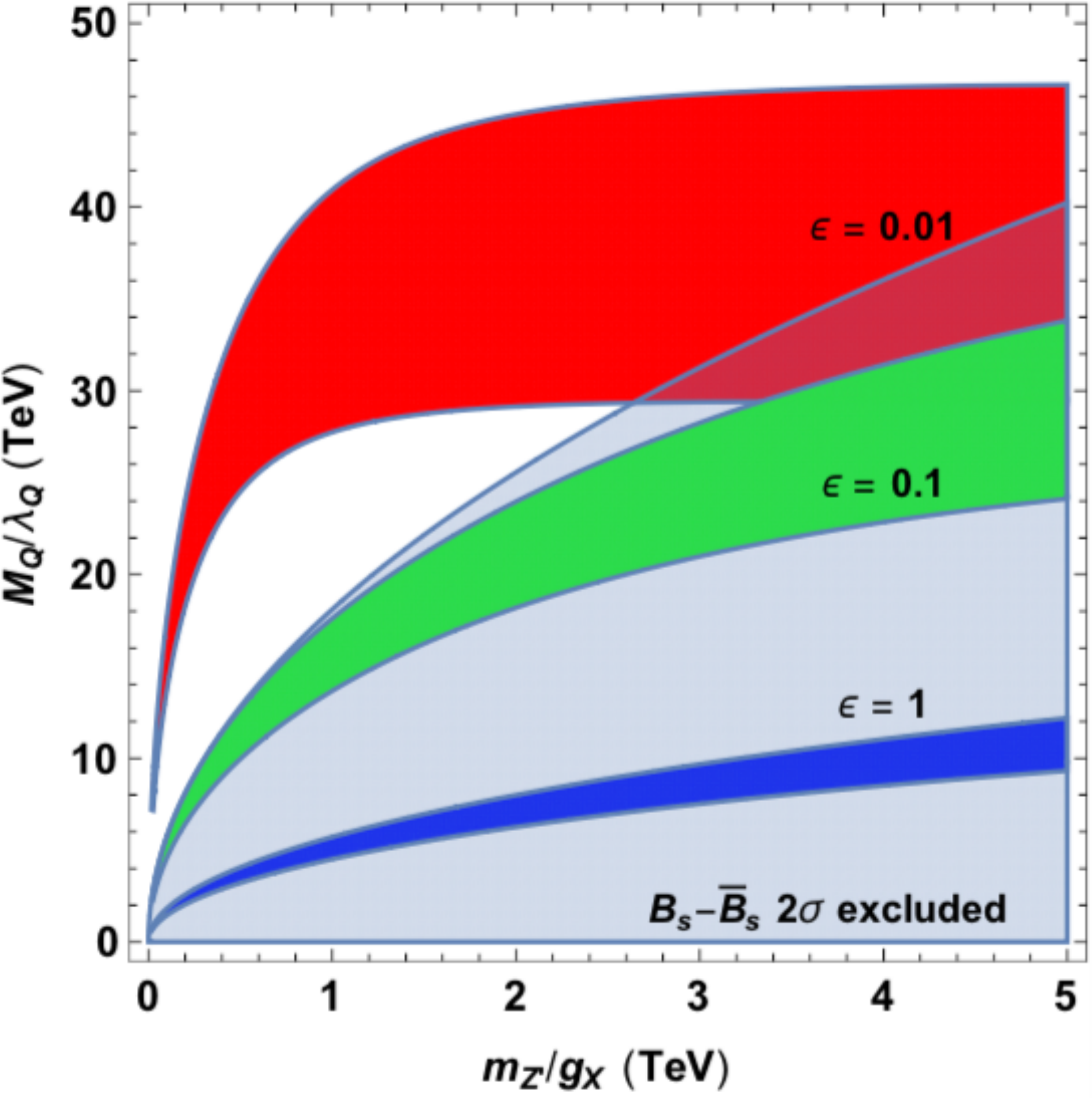}}

}%
\caption{\footnotesize Scan results for various models. 
\label{fig:mod}}
\end{figure}
We present the marginalized 2-dimensional posterior pdf in the ($m_{Z'}/g_X$, $M_Q/\lambda_Q$) plane in Model~2 in left of Figure \ref{fig:mod}. The VL mass range lies around a 20--30\tev\ scale for a coupling $\lambda_Q$ of order unity whereas the $m_{Z'}/g_X$ mass is limited to values below 5\tev, as a result of the $B_s$ mixing constraint. We find from middle of Figure \ref{fig:mod} that in both Model~1 and Model~2, the second VL mass is unbounded from above at the $2\,\sigma$ level. This is a consequence of the fact that $C_{9,\textrm{NP}}^{\prime \mu}$ in Model~1 and, especially $C_{10,\textrm{NP}}^{\mu}$ in Model~2, are consistent with zero at the $2\,\sigma$ level.\\
The $2\,\sigma$ regions of the 1-dimensional fit read 
\be
C_9^{\mu}=-C_{10}^{\mu}\in\left(-0.68,-0.29\right)\label{2sigma1dim}\,.
\ee
We apply this bound with $B_s$-mixing and show the favored $2\sigma$ region with different value of the hierarchical parameter $\epsilon$, defined as
$M_L/\lambda_{L,2}=\epsilon M_Q/\lambda_Q$ in right of Figure \ref{fig:mod}.

\subsection{Leptoquark}
Leptoquarks are considered potential candidate to explain the present flavor physics data. We consider a scalar leptoquark $S_3$ which Lagrangian acquires a Yukawa term
\be
\mathcal{L}\supset Y_{ij}Q^T_i(i\sigma_2)S_3 L_j+\textrm{H.c.}\,,
\ee
The tree level contribution is 
\be
C_9^{\mu}=-C_{10}^{\mu}=\frac{\pi v^2}{V_{tb}V_{ts}^{\ast}\,\alpha_{\textrm{em}}}\frac{\hat{Y}_{b\mu}\hat{Y}_{s\mu}^{\ast}}{m_{S_3}^2}\,.
\ee

The constraint from the 1-dimensional EFT at $2\,\sigma$ is given in \ref{2sigma1dim}.
This leads to
\be\label{lepto}
0.4\times 10^{-3} \left(\frac{m_{S_3}}{\tev}\right)^2 \leq\hat{Y}_{b\mu}\hat{Y}_{s\mu}^{\ast}\leq 1.1\times 10^{-3} \left(\frac{m_{S_3}}{\tev}\right)^2\,.
\ee

The most dangerous constraint is possibly given by $B\to K^{(\ast)}\nu\bar{\nu}$ decay. We get the limit
\be
\Re(\hat{Y}_{b\mu}\hat{Y}_{s\mu}^{\ast})\lesssim 2.2\times 10^{-2}\left(\frac{m_{S_3}}{\tev}\right)^2\,,
\ee
which does not constrain the parameter space emerging in Eq. (\ref{lepto}).

\begin{acknowledgments}
KK and DK are supported in part by the National Science Centre (Poland) under the research Grant No.~2017/26/E/ST2/00470.
EMS is supported in part by the National Science Centre (Poland) under the research Grant No.~2017/26/D/ST2/00490. 
The use of the CIS computer cluster at the National Centre for Nuclear Research in Warsaw is gratefully acknowledged.
\end{acknowledgments}

\bigskip 

\end{document}